_______________________________________________________________

# CAN SMALL MUSEUMS DEVELOP COMPELLING, EDUCATIONAL AND ACCESSIBLE WEB RESOURCES? THE CASE OF ACCADEMIA CARRARA

**Silvia Filippini-Fantoni**
University of Paris I – Sorbonne, Centre de Recherche Images et Cognitions, France
Silvia.Filippini-Fantoni@malix.univ-paris1.fr
http://www.univ-paris1.fr

**Jonathan P. Bowen**
London South Bank University, Institute for Computing Research
Faculty of BCIM, Borough Road, London SE1 0AA, UK
jonathan.bowen@lsbu.ac.uk
http://www.lsbu.ac.uk/bcimicr & http://www.jpbowen.com

**Abstract**
**Due to the lack of budget, competence, personnel and time, small museums are often unable to develop compelling, educational and accessible web resources for their permanent collections or temporary exhibitions. In an attempt to prove that investing in these types of resources can be very fruitful even for small institutions, we will illustrate the case of Accademia Carrara, a museum in Bergamo, northern Italy, which, for a current temporary exhibition on Cézanne and Renoir's masterpieces from the Paul Guillaume collection, developed a series of multimedia applications, including an accessible website, rich in content and educational material [www.cezannerenoir.it].**

## SMALL MUSEUMS AND THE DIFFICULTIES OF DEVELOPING MULTIMEDIA RESOURCES

Developing compelling educational and accessible web resources for the public often demands time, money and competences that most cultural institutions, museums in particular, do not necessarily possess. The problem is even more evident in small institutions (e.g., see [5]) where most of the energies are invested in what museum administrators not unreasonably consider more urgent needs such as security, personnel, restoration, temporary exhibitions, leaving the "educational" and "communicational" aspects, including the web, as a secondary issue.

One of the factor that has contributed to this situation is the lack of financial support from local or national government (especially in southern European countries) which, after having contributed to the collections' digitalization process, are now somewhat less eager to invest in the transformation of these resources into educational and entertaining services for the public. This means that cultural institutions have to finance these projects within heir own budget, leaving little margin for manoeuvre by small museums that already have difficulty surviving on their regular budgets.

Even when external funding is made available for the development of multimedia resources, through the support of private sponsors, the projects are usually carried out outside of the institution itself by external companies. There is no real integration of the project into the museum structure and therefore there is no development of internal competence that could be really beneficial in an environment where the already scarce personnel have little expertise in this respect.

It is important to realize that in the majority of small museums, the personnel is made up of volunteers or underpaid people who lack not only the competence and the motivation to encourage the development of multimedia and web resources but also the vision to develop a long term plan for the introduction and use of these resources in the museum on an ongoing basis.

Another major factor that acts against the introduction and development of multimedia resources in small local museums is the lack of understanding of technology by the public. This problem is even more evident in south and east parts of Europe where, as a result of the "digital divide", many categories of potential museum visitors such as the elderly, teachers and even young people do not have real access to the web and therefore do not feel comfortable with technology.

As a result of these factors very few small museums have managed to develop compelling, educational and accessible web resources for their permanent collections or temporary exhibitions that facilitate access to the museums' contents by visitors.

## THE CASE OF ACCADEMIA CARRARA

A typical example of the situation described above is represented by Pinacoteca Carrara, a small medieval and renaissance art museum located in the centre of Bergamo, northern Italy. The museum's collection, which comprises more than 20,000 pieces, (of which only less then 500 are exhibited) including some important works by Mantegna, Raphael, Botticelli and Pisanello, is the result of donations from local families and well-known people including Giacomo Carrara, an 18[th] century art lover and collector, to whom the museum is dedicated. The museum was established in 1810 and was run by a private foundation until 1958. Since then it has been administered by the local municipality which, after a long period of stagnation during the 1970s and 1980s, has decided to revive the image of the institution by organizing a series of major exhibitions on important artists such as Caravaggio, Lorenzo Lotto (both active in the area) and more recently Cézanne and Renoir. These exhibitions have brought very many visitors (more than 140,000 per exhibition) to the museum and have contributed to give visibility to a beautiful if small collection that would otherwise remain largely unknown to the general public.

Despite these huge successes in terms of public and visibility, the reality is still that of a small museum. Pinacoteca Carrara still remains a small institution. It is run by three employees, a Director, a few guards, and a group of freelance guides responsible for the guided tours and some educational initiatives. The museum is visited on average by 2,000 people per month (except for the periods in which major temporary exhibitions are on show, during which the number of visitors grows hugely). However it lacks most of the services and facilities that are normally considered of fundamental importance for a museum, such as an education, communication or marketing officer/department.

However, its new young director, Giovanni Valagussa, through his experience as an art historian and museology professor at the University of Brescia, has become aware of the importance of these services for the museum and the community and is therefore trying to raise some awareness of the need for more funding to be invested in this area. In particular, he recognizes the value of technology and multimedia applications as an important means of communication and education for the public. Therefore, in

___________________________________________________________________

cooperation with COBE, the company set up by the local administration to help the museum organize its temporary exhibitions, he has managed to find private sponsors to finance a few initiatives in this direction.

## THE CÉZANNE–RENOIR EXHIBITION

The first of these initiatives, which has included the launch of a new museum website and the further digitization of the collection, has been developed in relation to the new temporary exhibition on "*Cézanne–Renoir: 30 Masterpieces from the Paul Guillaume's collection*", which opened on 22 March 2005 in the museum's exhibition rooms, on show until 3 July 2005.

This exhibition presents, for the first time in Italy, 32 paintings, forming a fundamental group in the collection of the **Musée de l'Orangerie,** Paris. In particular, its structure is so devised as to offer visitors the necessary tools to understand more in depth the historical and artistic importance of the **Impressionist phenomenon** and its major role as a bridge with modern art. The exhibition focuses on the two opposite and complementary personalities of **Paul Cézanne** and **Pierre-Auguste Renoir**, who represent in a sense the two poles of this movement**.** Emphasis is placed also on the cultural importance of the collecting phenomenon, in this case represented by **Paul Guillaume** (1891–1934), an art critic, art dealer and patron based in Montparnasse, Paris whose selected paintings, together with those of his wife, Domenica Walter, built the **Musée de l'Orangerie**, identifying Cézanne and Renoir as the reference models for the young artists of the new generation, including Modigliani and Soutine.

In order to help in illustrating these themes and facilitate their understanding by the public, the museum has included the following in the exhibition space:

- Description panels for each exhibition room.
- MP3 disk drive random access audio guides that give the visitor the opportunity to listen to explanations and comments on the main themes of the different rooms, on the life and artistic activity of the main subjects of the exhibition (**Cézanne, Renoir and Paul Guillaume**), as well as on each of the exhibited paintings.
- A series of multimedia applications.

The multimedia applications, developed by the architect Paolo Venier – PPV, Milan, in cooperation with COBE Direzionale SpA, Bergamo and the museum, include:

**The palette**: Two monitors are installed at the entries of the second floor hallway. These monitors show some of the paintings by **Cézanne** and **Renoir**. Three pointers move on these virtual paintings at random. Each pointer corresponds to one of the three light sources that illuminate the ceiling vault. While moving at random on the surface of the painting, each pointer goes through a different colour area. The colour that each time is reached by the pointer is instantaneously sampled and sent to the video-projector. Transformed into a light beam, the sampled colour is projected on the ceiling vault. Visitors find themselves plunged in colours, as if they were actually going through **the colour palette** used by Cézanne or Renoir.

**The videos**: Some rare documentaries and films on **Cézanne** and **Renoir** are shown. Through testimonies and critical analysis of their work, the artists are identified as two of the major influences in the artistic revolution that occurred in early 20[th] century.

**The virtual book**: The opinions of several art critics and historians who, over time, have expressed their views on the work of these painters are collected on a wall monitor. These opinions can be turned over on the screen, like the pages of a virtual book.

## THE WEBSITE

Besides integrating technology into the exhibits to support the "educational and entertaining" experience of the visitors, the museum has invested in the creation of a website. The facility [www.cezannerenoir.it] was developed in less than two months through cooperation between the museum, the University of Paris multimedia design programme, and two local companies (ICTeam [www.icteam.it] for the technical aspects and BlastMedia [www.blastmedia.it] for the graphic design and the accessibility issues). Despite being quite traditional in its approach, the website offers a compelling and rich "edutaining" experience for different types of visitors (tourists, children, teachers, families, experts).

The focus on education and entertainment has been our priority in setting up the site because school groups, which represent the most important categories of visitor to the Pinacoteca, can strongly benefit from a website that offers different types of educational material aimed at supporting and preparing for an actual visit. Moreover, the need to attract new kinds of audience, such as young people and families, who are usually not frequent visitors to the exhibitions organized by Pinacoteca Carrara, was a further motivation for focusing on educational aspects.

Besides trying to answer the needs of these specific categories of users, the exhibition website also aims at:

1) facilitating everybody's access to the exhibition's main themes through an exhaustive presentation of its content;
2) encouraging actual attendance by providing information and services that facilitate access to the museum;
3) providing information about parallel events and initiatives;
4) guaranteeing good design for usability and accessibility.

Keeping this in mind, we organized the website in six different sections, some of which are merely informative (Information, Initiatives, L'800 della Carrara) while others provide more educational and entertaining tools and applications developed for different users groups (Games, Education and The exhibition). Here is a short description of the different sections as well as a screenshot from the website's home page (Figure 1):

**Information**: in this section users can find information on how to reach the museum (address, exhibition dates, ticket price, ticket booking, etc.) and on visiting the exhibition (guided visits, audio guides);

**The exhibition:** where users can find an exhaustive presentation of the exhibition themes (exhibited works, exhibition rooms, central characters), as well as some introductory comments on the Musée de l'Orangerie in Paris;

**Games:** where children have the opportunity to interactively explore some selected exhibited works and take part in some entertaining educational games;

**Education:** conceived especially for parents and teachers looking for educational material aimed at supporting and preparing for a visit.

**Events**: users can find in this section some useful information on events and initiatives connected with the exhibition, organized in the city of Bergamo and its province (lectures, films, etc.);

**L'800 della Carrara** (The 19[th] century at Accademia Carrara): In this section users can find an introduction to the exhibition on the lesser-known 19[th] century works owned by the Accademia Carrara, held during the same period in the permanent collection hall.

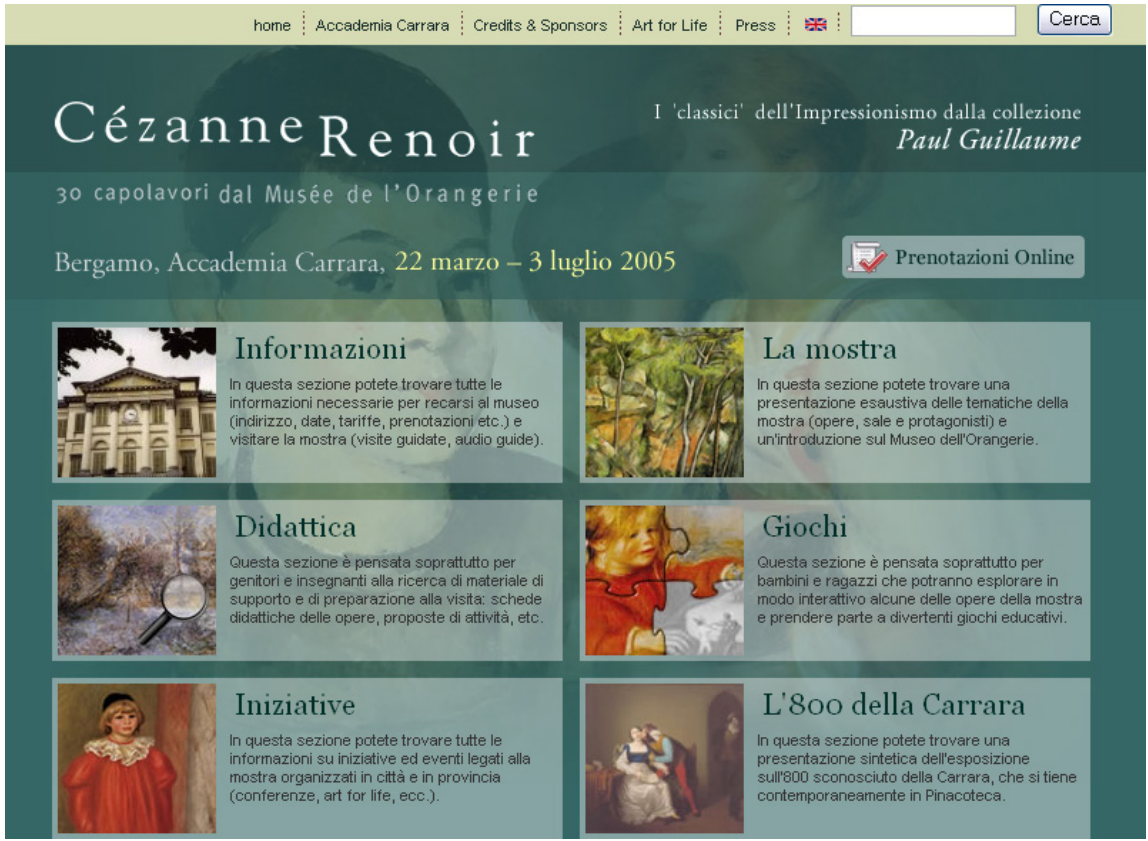

**Figure 1:** Cézanne–Renoir website home page [www.cezannerenoir.it].

## THE PROJECT'S EDUCATIONAL BACKGROUND

Given the strong focus on education, we decided that, for the development of the website, especially its educational content and applications, we would draw inspiration from the available literature on multimedia and learning. In particular, we chose to refer to the research in [11] which, taking into consideration the theories of David Kolb and Howard Gardner on the existence of different learning styles and levels of intelligence, identifies six types of possible web-based learning activities. These types of activity, as described in [11], are:

- *Creative Play:* emphasizes both open-ended self-expression and the application of subject knowledge, together with concepts to aid in building some kind of product such as a story, picture, movie, etc.

- *Guided Tour:* proposes the exploration of a topic chosen by an expert. The guide leads the users on their path through the topic.

- ***Interactive Reference:*** allows the user to explore a topic of personal interest through informative words and pictures and by following the links that are of most interest.

- ***Puzzle/Interactive Mystery:*** involves analysis and deductive or inductive reasoning to reach a logical solution. The user relies on evidence from people, nature, or reference material provided by the activity to solve the problem.

- ***Role-playing Story:*** allows users to adopt a persona different from their own, giving them the ability to do things they cannot ordinarily do (e.g., break natural or societal laws, experience people and places that are normally out of reach). They can also interact with other characters, whose behaviour either may be scripted or controlled by other players.

- ***Simulation:*** allows the user to run a model of the real world and see what happens when they change something. The choices the user makes determine the results.

In order to meet the different needs of all possible users, we tried to propose learning solutions that would take into consideration some or a combination of the six approaches above. Moreover, in the development of the different "educating" applications for the website, we aimed to apply the concept that "there are clear differences in the type of web-based learning activity that adults prefer in comparison to children" [11]. Adults prefer the information-based activities of ***Interactive Reference*** and ***Simulation***, whereas children, not surprisingly, are more inclined to prefer the exploratory experiences of ***Role-playing Story***, ***Creative Play*** and ***Puzzle/Mystery***. Such differences are motivated in [11] by the fact that adults apparently bring an intrinsic motivation to the learning experience; they know what they want to learn and they want to learn it in the most direct way. Children, on the other hand, need to be motivated. They respond positively to the opportunity for interaction and choice within a goal-based environment that offers them an extrinsic purpose.

These important theories and differences were particularly taken into consideration for the definition, creation and setting up of the sections of the sites with the strongest educational value, in which the exhibitions content and themes where presented (i.e., covering the exhibition, games and education). A short description of these three different website sections and the principles applied is given in the sections below.

## THE EXHIBITION

***Interactive Reference*** is the dominating principle applied in this section, meant especially for those adult users that are interested in exploring the main themes of the exhibition in some depth. The exhibition contents (introduction, information about Cézanne, Renoir and Paul Guillaume, descriptions of the art works, etc.) are presented here in such a way as to allow the user to explore the subject of most interest in an independent way, navigating from one section to another by simply following the interactive links on the pages. A further possibility to explore subjects of interest in a similarly interactive way is also offered in the "education" section (see later for more details) where a table with a chronology of events, a list of references and links to external websites on the two artists are made available for those who wish to research the exhibition's themes even further.

Of particular interest from the point of view of the ***Interactive Reference*** model is the presentation of the 32 exhibited artworks (see Figure 2, for example) which, besides including a short textual description, enhanced with interactive links to other pages, also gives the possibility of the following activities:

1) **Exploring the painting in detail** through a device that works like Microsoft Magnifier (see Figure 3); by moving a lens on the painting it is possible to observe the details in another frame (available for all paintings, but only for Internet Explorer);

2) Listening to the **audio description** of the artwork, extracted from the audio guide available in the museum (available only for a select number of paintings);

3) Accessing the **educational card** (see the EDUCATION section later for more details), available in PDF (Portable Document Format), which provides an analysis of the exhibited work conceived for children of a certain age group (offered for only a selection of the paintings);

4) Viewing the **interactive description** of the artwork, themes and techniques used by the artist (see the GAMES section later for more details) conceived specifically for children of different age groups (available only for a selected number of paintings);

5) Accessing a list of **activity proposals** for  children of different ages inspired by the artwork in question;

6) **Exploring the relation between two or more exhibited works**, as in the real exhibition, where some of the paintings are compared to underline differences or similarities between the two artists, or between the different phases of their artistic evolution.

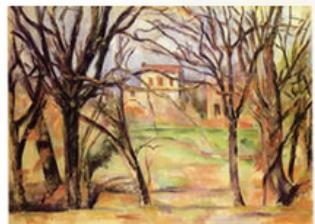

**Figure 2:** Example of a picture in the exhibition.

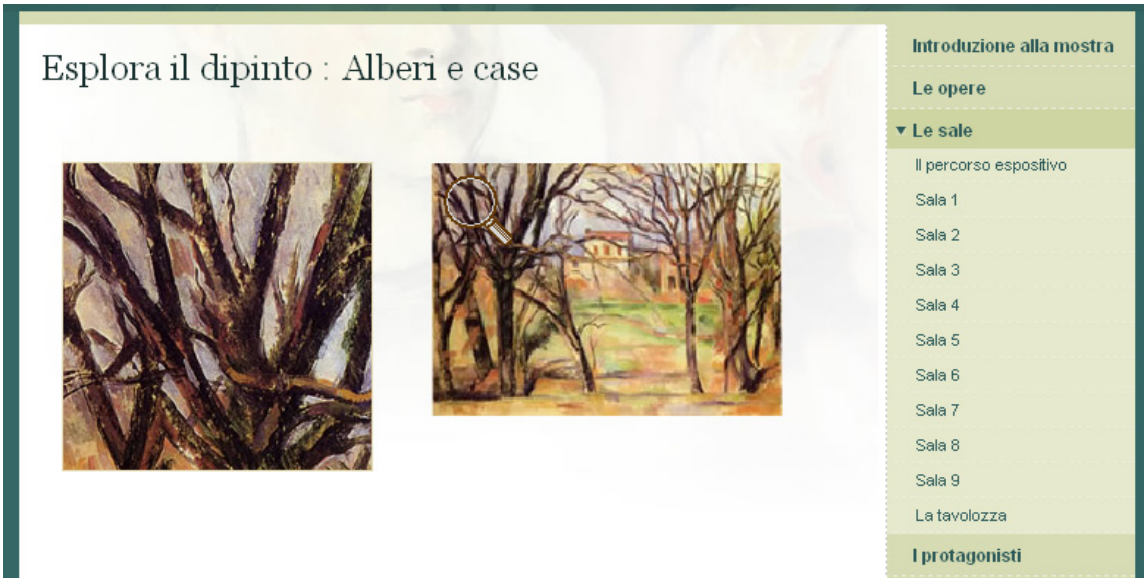

**Figure 3:** Exploring the detail of a painting.

Besides the interactive reference model, in this section it is possible to find an example of a **guided tour** (see Figure 4). A description of the exhibition structure, presenting the main themes of each room and a brief introduction to the exhibited artwork is available for those who feel less comfortable about exploring the subjects independently and would rather be guided step by step through the exhibition contents, as conceived by the curators. However, these descriptions are also enriched with a series of interactive links to other pages of the website for those who, after the guided tour, wish to continue the exploration in an independent and non-sequential manner.

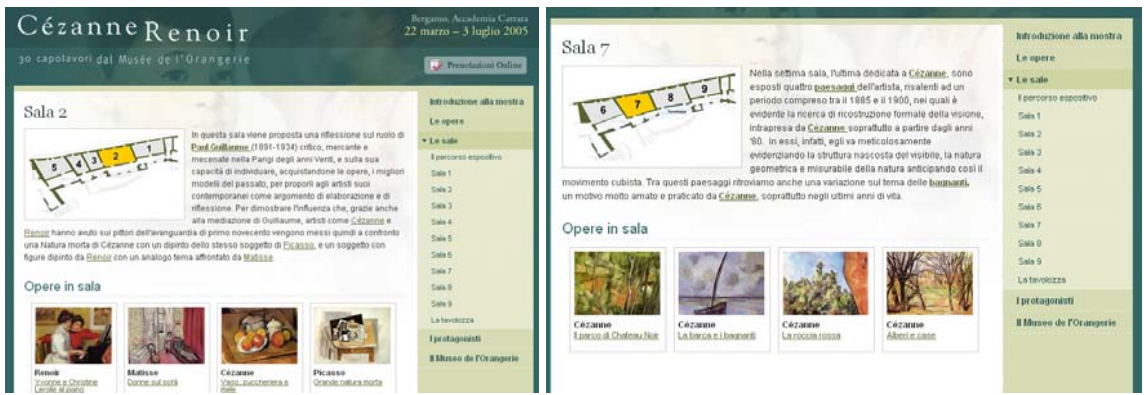

**Figure 4:** Guided tour, exploring rooms in the exhibition.

## GAMES

This section (**available only in Italian**) has been conceived especially to facilitate children's access to the exhibition contents and themes, for whom an exploration through the interactive reference model that was applied in "The exhibition" website section is not suitable (see the project's educational background for more detail). The types of web-based learning activities proposed here are inspired by the "quiz", "mystery" and "role play" approaches, which we have seen are preferred by children because they give more "opportunity for interaction and choice […] that offers them an extrinsic purpose" [11].

The **interactive description** of some of the artworks (see Figure 5), for example, is organized around a series of questions concerning the painting that the users, in this case, children of different age groups, have to answer by choosing among a series of three possibilities. In such a way children can become aware of certain details of the artwork that become the starting point to explain the techniques used by the artists, the subjects of the artworks, the artistic movements to which they belong, the artists' evolution, or the works of art that inspired them. All this is done by using a very simple language adapted to the age level for which the interactive description has been conceived. To reinstate some of the explored concepts we also included, whenever available, some texts, narrated by an actor, extracted from Cézanne and Renoir's writings in which the same idea was expressed through the artists' words. The paintings for which this type of analysis (questions/explanation/text) is proposed are *four works* by **Pierre-Auguste Renoir**, along with some *landscapes* and *still life* compositions by **Paul Cézanne**. The reference models for this type of activity are both the quiz (questions) and the guided tour as the user is guided step by step through the analysis of the artwork by a kind of virtual teacher/curator.

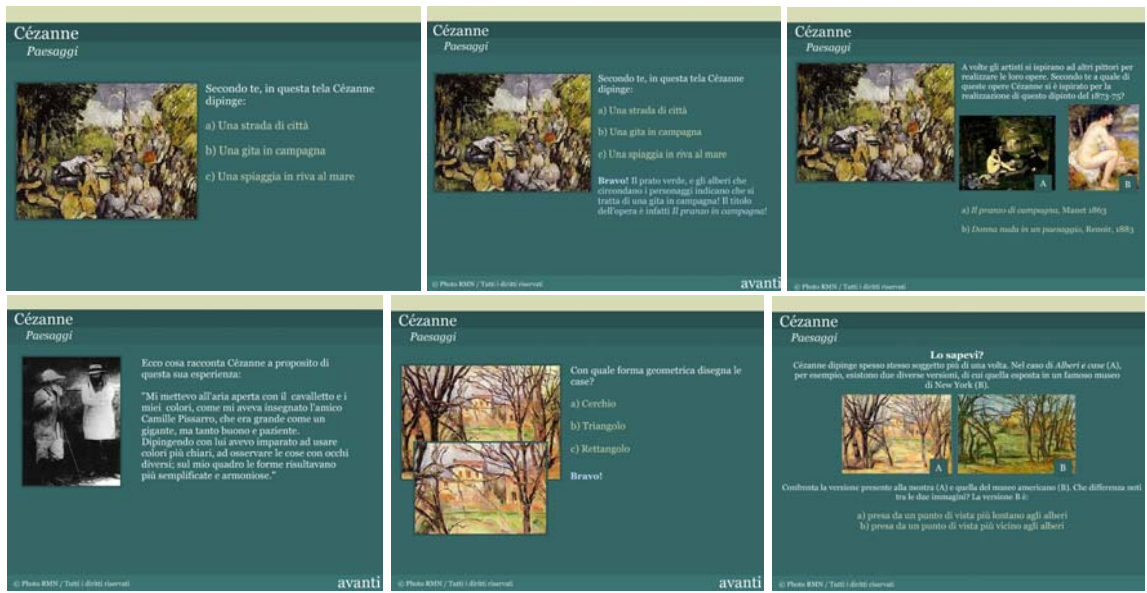

**Figure 5:** Interactive description of pictures.

Besides these interactive analyses of some of the artworks, this section also offers a challenging and entertaining educational game (see Figure 6): *Catch Arsène Lupin!* This interactive game, inspired by the imaginary character of the thief Arsène Lupin, who is familiar to Italian children thanks to the Japanese cartoon series, the reruns of which are still broadcast on TV, has been conceived for children of age 9 and older. The aim of the game is to help Inspector Zenigatta of the French police to find out, through a series of clues that are provided only to those who correctly answer some questions concerning the life and works for Cézanne and Renoir, where Lupin is hiding and to recover the painting stolen by him back to the collector Paul Guillaume. The game offers the possibility not only to test their knowledge acquired by the children during a visit to the exhibition (or navigation through the website), but also to find out more about some famous Parisian locations familiar to Guillaume, Cézanne and Renoir. The types of web-based learning activities proposed here are: **"role play"** (children have to behave like a Inspector in the French police and pursue the thief) and **"quiz/mystery"** (where the user relies on evidence from people, or reference material provided by the activity to find a solution to the problem). Both stimulate the child towards a greater emotional involvement in the activity, aimed at facilitating the learning process.

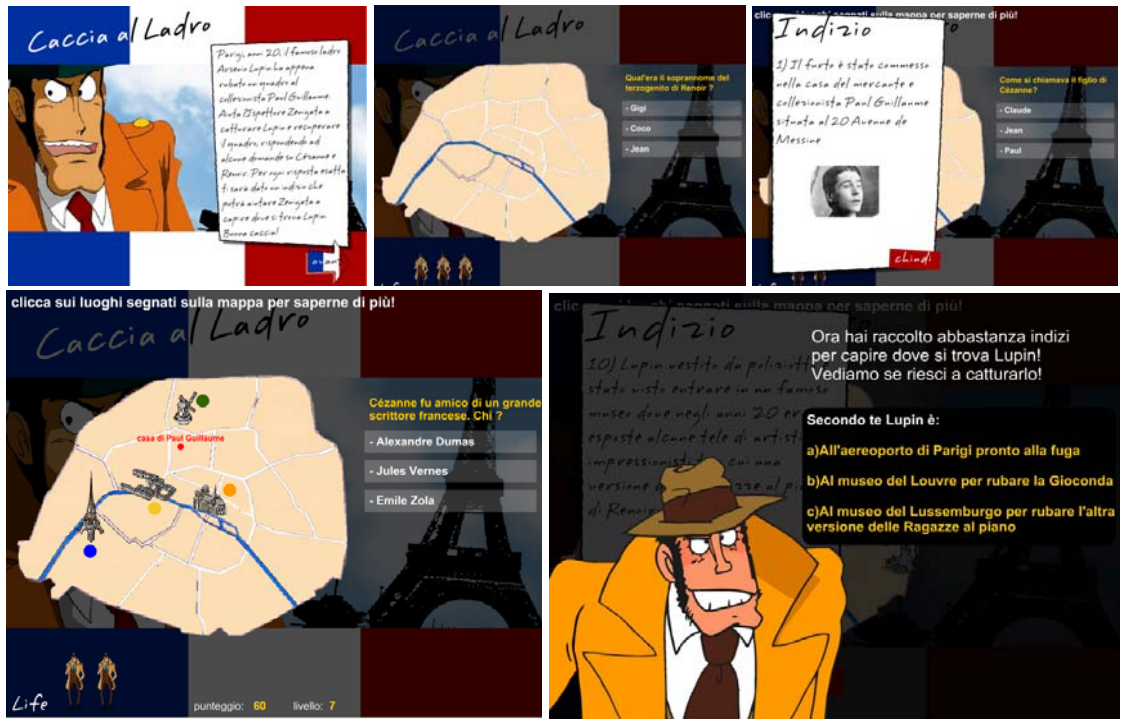

**Figure 6:** Interactive educational game: *Catch Arsène Lupin!*

If the game was to be redesigned in the future, or a new one was to be added, more account could be taken of difference between boys and girls, for example, when playing online games since they may have different goals in their role-playing [1].

## EDUCATION

This section (**available only in Italian**) is particularly aimed at parents and teachers looking for educational material aimed at supporting and preparing for a visit to the exhibition. It includes:

a) **Educational cards** (see also the Exhibition section above) available in PDF format, providing an analysis of the exhibited artworks conceived for children of different age groups (from 5 to 18). These cards can be used directly by the children or even by their parents or teachers as a starting point for an analysis of the artworks to be carried out at home or in the classroom before, during or after an actual visit.

b) A list of **activity proposals** (inspired by the artworks, their subject or techniques), to be carried out before, during or after the visit, conceived for children of different age groups. For the proposed activities, we referred to and drew inspiration from the same models proposed in [11] for the interactive learning activities: **creative play** (such as drawing and decoupage), **role play**, but also (for older children) more information-based types of activity such as research proposals. The children's **questionnaire/guide** supporting the exhibition, available in PDF format,

___________________________________________________________________

is of particular note. Besides including a brief description of the rooms and some interesting information about the exhibited artworks, children can find a list of questions to which they have to look for the answers in the exhibition itself.

c) A table with a **chronology of the events** related to the themes of the two exhibitions on show (L'800 della Carrara and Cézanne–Renoir), enriched with links to other sections of the site.

d) A list of references (**bibliography**) on the two artists.

e) A **list of websites** on the two artists.

The idea behind this section is to offer a real educational platform of material (downloadable for later use and non-downloadable for immediate use) for teachers and students, to be used in the museum, in the classroom or at home, before, during or after the visit. Moreover, for those teachers or students that are interested in exploring in more detail the themes related to the exhibition, extra information is available in the form of a list of references, a list of websites and a chronology of events.

Educational aspects were considered to be a very important issue for the website. Also of considerable and increasing related concern is the issue of usability in general and accessibility in particular. There are moral and legal reasons to make websites accessible to as many people as possible, whether they are children, disabled, the elderly, etc. All types of cultural institutions, both large and small, should be considering accessibility very seriously [3],[4], as covered further in the next section.

## ACCESSIBILITY

Accessibility is a very important but delicate issue that is often put aside by museums in general and small institutions in particular, due to the lack of funds and competence or interest of staff. This problem is evident in Italy, despite the existence of *Law n. 4, January 9, 2004* [9]. There are provisions to support access to information technologies for the disabled (also known as "The Stanca Act"), which obliges public organizations to have accessible websites, but the actual initiatives in this respect are still very limited. Local and national museums under public administration are obliged to conform to this legislation, which indicates February 2006 as the ultimate date for adaptation. The law, which is in line with the *Web Content Accessibility Guidelines* (WCAG) Version 1.0 guidelines [7], introduces two verification levels: a technical one (following 22 control rules) and a subjective one (to be carried out by a third person, possibly disabled, according to empirical criteria and evaluation).

As far as the website [www.cezannerenoir.it] is concerned, although we were still not "obliged" to provide accessible solutions, we thought, working with ICTeam and BlastMedia, two companies that have much experience with accessibility, it would be worth aiming to comply with the legislation now, in an attempt to gain experience of the accessibility issue for the necessary update of the museum's official website, which is due to take place in 2006.

Keeping this in mind, we developed a site suitable to be used and explored by several different types of user, including people with disabilities. From a technical point of view, we have aimed at:

a) ensuring safe navigation with any browser by following World Wide Web Consortium (W3C) web page markup standards (specifically XHTML 1.0

_______________________________________________________________________

[13]) – validation can be performed online through a W3C web facility [validator.w3.org];

b) leaving the task of managing the graphical layout to Cascading Style Sheets (CCS) [12];

c) identifying the exceptions to the natural language of the site: commenting links, describing images, incorporating PDF documents;

d) allowing each main navigation link also to be activated through the keyboard as well as the mouse.

In some cases, we had to resort to the use of technical solutions that are sometimes seen as less accessible, such as equipping written explanations with animations or audio-files programmed with Macromedia Flash, in order to "animate" the context, but these are secondary elements.

In practical terms, this site is:

**1)** navigable via a keyboard;

**2)** accessible from an audio browser or screen reader (such as the widely-used JAWS software [www.freedomscientific.com]);

**3)** accessible using a text-only browser (such as Lynx [lynx.browser.org]).

The site offers users the opportunity to readjust the text by changing the format in the menu of their own browser. It may also be viewed in a customizable text-only form [access.museophile.net/www.cezannerenoir.it] using the *Betsie* open source software, originally developed by the BBC [betsie.sourceforge.net].

All these expedients make the Cézanne–Renoir website one of the most accessible museum websites in Italy. Considering that this was achieved by a small museum with very little resources demonstrates that accessibility is possible even when budgets are tight if considered from the start during the design. For success, it is necessary that both the museum and the designers consider it an important issue. Extra financial investment is a lesser concern in practice.

Note that semi-automatic checking for web page accessibility conformance and other related issues such as quality and privacy can be undertaken online using Watchfire's excellent newly established WebXACT website [webxact.watchfire.com].

## CONCLUSION: THE RESULTS

During its first month of activity, the website registered more than 30,000 virtual visitors (about the same as in the real exhibition, which has also been a success), 274,000 visited pages, an average of eight pages per visit viewed, and a high number of repeated visitors (53%). If we look at the contents, the most accessed pages are: the home page and the online reservation system, followed by some distance by the description of the art works and the games. Among the PDF files, the most downloaded were the educational cards, followed by the guide/questionnaire for the exhibition, with a total of almost 10,000 downloads. Even the MP3 files, with audio description of some of the artworks have been, quite successful (with around 7,000 downloads).

Despite the limited affirmation of the use of new technologies in Italian schools, due especially to the teachers' uneasiness with technology and the Internet, the high number of downloaded documents and the success of the games are encouraging. Further research needs to be done to test whether the website contributed effectively to the success of the exhibition. Certainly compared to last year's exhibition on the portrait artist Fra Galgario, when the associated website offered fewer services and less educational content, the results are quite impressive.

The success of the site has paved the way for a future collaboration between the museum, the university and the two companies involved, for the reorganization of the museums' official website, planned for 2006. For this project, we will try to involve the museum's personnel to a greater extent, especially with regard to the creation of content and the editorial process, in such a way as to allow the museum to be able to independently update the system alone. This will help to create new expertise within the museum and hopefully will contribute to the definition of a long-term plan for the creation of multimedia resources aimed at the public. This will be especially helpful in view of the planned physical renovation of the museum over the coming years.

This website has proved that investing in the development of compelling and educational multimedia resources can be very fruitful, especially for small museums that can gain not only greater visibility, but also in terms of competence and experience. Raising the online profile of a museum is important, but may be especially difficult for a lesser-known museum largely using a non-English language on its website [10]. For the future, smaller museums may wish to consider additional aspects like website personalization [6],[8] and the development of online communities [2], which are currently mainly of concern to larger museums.

## ACKNOWLEDGEMENT

Silvia Filippini-Fantoni is a Visiting Research Fellow in the Institute for Computing Research at London South Bank University from June 2005 to August 2005. Attendance at the *EVA London 2005 Conference* has been part-funded by London South Bank University.

_______________________________________________________________